\documentclass[12pt]{article}
\usepackage{a4}
\usepackage{amsfonts,amssymb,amsmath}
\usepackage{graphicx}
\usepackage{cite}

\begin{document}

\title{Hydrogen atom on curved noncommutative space}
\author{V.G. Kupriyanov\thanks{%
e-mail: vladislav.kupriyanov@gmail.com} \\
CMCC, Universidade Federal do ABC, Santo Andr\'{e}, SP, Brazil}
\date{\today                                        }
\maketitle

\begin{abstract}
We have calculated the hydrogen atom spectrum on curved noncommutative space
defined by the commutation relations $\left[ \hat {x}^{i},\hat{x}^{j}\right]
=i\theta\hat{\omega}^{ij}\left( \hat {x}\right) $, where $\theta$ is the
parameter of noncommutativity. The external antisymmetric field which
determines the noncommutativity is chosen as $\omega^{ij}(x)
=\varepsilon^{ijk}{x}_{k}f\left( {x_i}x^{i}\right) $. In this case the
rotational symmetry of the system is conserved, preserving the degeneracy of
the energy spectrum. The contribution of the noncommutativity appears as a
correction to the fine structure. The corresponding nonlocality is
calculated: $\Delta x\Delta y \geq \frac{\theta^2}{4} |m\langle f^2\rangle| $,
where $m$ is a magnetic quantum number.
\end{abstract}

\section{Introduction}

A combination of quantum mechanical arguments with general relativity
indicates that at the distances of the order of the Planck length the
space-time is nonlocal, therefore cannot be described as a differentiable
manifold and should be treated as a kind of noncommutative structure \cite%
{Doplicher}. Consider the noncommutative space realized by the coordinate
operators $\hat{x}^{i},\,i=1,...,N,$ satisfying the algebra
\begin{equation}
\left[ \hat{x}^{i},\hat{x}^{j}\right] =i\theta\hat{\omega}^{ij}\left( \hat{x}%
\right) ,  \label{1}
\end{equation}
where $\hat{\omega}^{ij}\left( \hat{x}\right) $  is an operator describing
the noncommutativity of the space, and $\theta$ is the parameter of noncommutativity. The consistency condition for the algebra
(\ref{1}) implies, see e.g. \cite{KV}, that the symbol $\omega_{q}^{ij}%
\left( x\right)$ of the operator $\hat{\omega}^{ij}\left( \hat{x}\right) $
should have a form:
\begin{equation}
\omega_{q}^{ij}\left( x\right) =\omega^{ij}\left( x\right) +\omega
_{co}^{ij}\left( x\right) ,
\end{equation}
where $\omega^{ij}\left( x\right) $ is Poisson bi-vector, i.e., obey the
Jacobi identity%
\begin{equation}
\omega^{il}\partial_{l}\omega^{jk}+\omega^{kl}\partial_{l}\omega^{ij}+%
\omega^{jl}\partial_{l}\omega^{ki}=0,  \label{2}
\end{equation}
and the term $\omega_{co}^{ij}\left( x\right) $ stands for non-Poisson
corrections to $\omega^{ij}\left( x\right) $ of higher order in $\theta$,
expressed in terms of $\omega^{ij}\left( x\right) $ and its derivatives,
which depend on specific ordering of the operator $\hat{\omega}^{ij}\left(
\hat{x}\right) $. So, in order to determine the noncommutativity (\ref{1})
we should define from physical considerations the antisymmetric field $%
\omega^{ij}\left( x\right) $ obeying the eq. (\ref{2}) and specify the
ordering. In what follows we treat $\omega^{ij}\left( x\right) $ as an
external field and we choose the symmetric Weyl ordering.

This article aims to illustrate the possible phenomenological consequences
of the presence of the noncommutativity of the type (\ref{1}), when the
commutator between coordinates is a function of these coordinates. As a
starting point to study noncommutativity of the general form we consider quantum mechanics
(QM). Two-dimensional model of position-dependent noncommutativity in QM was
proposed in \cite{GK}. The particular example was considered in \cite{Fring1}, where also was observed that canonical operators in these models are in general non Hermitian with respect to standard inner products \cite{Fling2}. For the example of QM and field theory on kappa-Minkowski space see e.g. \cite{Meljanac} and references therein. Hydrogen atom in fuzzy spaces was discussed in \cite{Presnajder}. General form of coordinate dependent noncommutatiivity in QM was considered in \cite{kup14}. Note that the quantum mechanical scale of energies is rather
different from the Planck scale, however some important properties like
preservation of symmetries and corresponding consequences can be studied
already in QM.

In particular,  it is well known fact that the canonical
noncommutativity, $\left[ \hat{x}^{i},\hat{x}^{j}\right] =i\theta ^{ij},$
where $\theta ^{ij}$ is an antisymmetric constant matrix, breaks the
rotational symmetry of the Hydrogen atom, which removes the degeneracy of the energy
levels \cite{Chaichian}. This fact leads to the bounds of noncommutativity in this
model. The same logic remains in the field theory, see e.g. \cite{AGSV} and references therein. We will show here that the noncommutativity can be introduced in a way to preserve the symmetries and the corresponding degeneracy. To this end in Section 2 we describe the model of QM with NC coordinates \cite {kup14} and construct an explicit form of the trace functional on the algebra of the star product corresponding to (\ref{1}). In Section 3 we consider a particle in a central potential on curved NC space, defined by the external antisymmetric field $\omega^{ij}(x)
=\varepsilon^{ijk}{x}_{k}f\left( {x_i}x^{i}\right) $. For simplicity we set ($m_e=c=\hbar =1
$).

\section{Quantum mechanics with noncommutative coordinates}

To define the QM on the noncommutative spaces of the type (\ref{1}) we will
need to introduce two objects, the star product and the trace functional.
The existence of the star product for any Poisson bi-vector $\omega
^{ij}\left( x\right) $ is guaranteed by the Formality Theorem by Kontsevich
\cite{Kontsevich}, the explicit construction of the star product up to the
fifth order in deformation parameter can be found in \cite{KV}. In
particular, for any two functions $f$ and $g$ it has a form
\begin{align}
& (f\star g)(x)=f\left( \hat{x}\right) g(x)=fg+\frac{i\theta }{2}\partial
_{i}f\omega ^{ij}\partial _{j}g  \label{star} \\
& -\frac{\theta ^{2}}{4}\left[ \frac{1}{2}\omega ^{ij}\omega ^{kl}\partial
_{i}\partial _{k}f\partial _{j}\partial _{l}g-\frac{1}{3}\omega
^{ij}\partial _{j}\omega ^{kl}\left( \partial _{i}\partial _{k}f\partial
_{l}g-\partial _{k}f\partial _{i}\partial _{l}g\right) \right] +O\left(
\theta ^{3}\right) .  \notag
\end{align}

The trace functional is determined as $Tr\left( f\right) =\int \mathbf{%
\Omega }\left( x\right) f\left( x\right) ,$ where $\mathbf{\Omega }\left(
x\right) $ is an integration measure. The trace should obey the property
\begin{equation}
Tr\left( f\star g\right) -Tr\left( fg\right) =0,  \label{trace}
\end{equation}%
which also guarantee the cyclic property of trace. The existence of a trace functional for
a Kontsevich star-product related to any Poisson bi-vector $\omega ^{ij}$ such that ${div}_\Omega\omega ^{ij}=0$ was demonstrated in \cite{FS}. But to make
calculations one needs an explicit form for it. Suppose that there exist a
function $\mu \left( x\right) $ such that
\begin{equation}
\partial _{i}\left( \mu \omega ^{ij}\right) =0.  \label{4}
\end{equation}%
Let $\mathbf{\Omega }\left( x\right) =d^{N}x\mu \left( x\right) ,$ so
\begin{equation}
Tr\left( f\right) =\int d^{N}x\mu \left( x\right) f\left( x\right) .
\label{5}
\end{equation}%
We should verify if this definition of trace is correct, i.e., if (\ref%
{trace}) holds true for any two functions $f$ and $g$ vanishing on the
infinity. In the zero order in $\theta $ eq. (\ref{trace}) is just an
identity, in the first order it is satisfied due to (\ref{4}). In the second
order in $\theta $ the right hand side of (\ref{trace}) is%
\begin{equation}
-\frac{\theta ^{2}}{4}\int d^{N}x\mu \left( x\right) \left[ \frac{1}{2}%
\omega ^{ij}\omega ^{kl}\partial _{i}\partial _{k}f\partial _{j}\partial
_{l}g-\frac{1}{3}\omega ^{ij}\partial _{j}\omega ^{kl}\left( \partial
_{i}\partial _{k}f\partial _{l}g-\partial _{k}f\partial _{i}\partial
_{l}g\right) \right]   \label{7}
\end{equation}%
Integrating this expression by parts on $f$ and $g$ and using (\ref{4}) we
rewrite it as%
\begin{equation}
\frac{\theta ^{2}}{12}\int d^{N}x\partial _{i}f\partial _{l}\left( \mu
\omega ^{ij}\partial _{j}\omega ^{lk}\right) \partial _{k}g,  \label{8}
\end{equation}%
where the matrix $\partial _{l}\left( \mu \omega ^{ij}\partial _{j}\omega
^{lk}\right) $ is symmetric, i.e.,
\begin{equation}
\partial _{l}\left( \mu \omega ^{ij}\partial _{j}\omega ^{lk}\right)
=\partial _{l}\left( \mu \omega ^{kj}\partial _{j}\omega ^{li}\right) ,
\label{9}
\end{equation}%
due to the Jacobi identity and (\ref{4}). The expression (\ref{8}) is
different from zero for two arbitrary functions $f$ and $g$. To solve this
problem we use the gauge freedom in the definition of the star product \cite%
{Kontsevich}. Let us construct a new star product
\begin{equation}
f\star ^{\prime }g=D^{-1}\left( Df\star Dg\right) ,  \label{10}
\end{equation}%
choosing a gauge operator $D$ in such a way that (\ref{trace}) holds true
for this new product. Taking into account (\ref{8}) and (\ref{9}) one should
looking for $D$ in the form%
\begin{equation}
D=1+\theta ^{2}b^{ik}\partial _{i}\partial _{k}+O\left( \theta ^{3}\right) .
\label{11}
\end{equation}%
In this case%
\begin{equation}
f\star ^{\prime }g=f\star g-2\theta ^{2}b^{ik}\partial _{i}f\partial
_{k}g+O\left( \theta ^{3}\right) .  \label{12}
\end{equation}%
The condition (\ref{trace}) for the star product (\ref{12}) in the second
order gives%
\begin{equation}
\frac{\theta ^{2}}{12}\partial _{i}f\partial _{l}\left( \mu \omega
^{ij}\partial _{j}\omega ^{lk}\right) \partial _{k}g-2\theta ^{2}\mu
b^{ik}\partial _{i}f\partial _{k}g=0.  \label{13}
\end{equation}%
That is,
\begin{equation}
b^{ik}=\frac{1}{24\mu }\partial _{l}\left( \mu \omega ^{ij}\partial
_{j}\omega ^{lk}\right) .  \label{14}
\end{equation}%
Note that the star product (\ref{12}) will be an associative if and only if $b^{ik}$ is a symmetric tensor, which in turn is provided by the eq. (\ref{9}). We conclude that given a Poisson bi-vector $\omega ^{ij}$ and a function $%
\mu \left( x\right) $ obeying (\ref{4}), the modified star product (\ref{12}) admits
the trace (\ref{5}).

Now, according to the \cite{kup14} we describe the model of QM on NC spaces of general form, defined by the algebra (\ref{1}). \textit{The Hilbert space} is determined as a space of
complex-valued functions which are square-integrable with a measure $\mathbf{%
\Omega }\left( x\right) $. \textit{The internal product} between two states $%
\varphi \left( x\right) $ and $\psi \left( x\right) $ from the Hilbert space
is defined as
\begin{equation}
\left\langle \varphi \right\vert \left. \psi \right\rangle =Tr\left( \varphi
^{\ast }\star^{\prime } \psi \right) .  \label{scalar}
\end{equation}%
\textit{The action of the coordinate operators $\hat{x}^{i}$} on functions $\psi (x)$ from
the Hilbert space is defined through the modified star product (\ref{12}), for any function $%
V\left( x\right) $ one has
\begin{equation}
{V}\left( \hat{x}\right) \psi (x)=V(x)\star^{\prime } \psi (x).  \label{15}
\end{equation}%
In particular, from $\hat x^i \psi=x^i\star^{\prime}\psi$, one may see that
\begin{equation}
\hat x^i=x^i+\frac{i\theta}{2}\omega^{ij}\partial_j-\frac{\theta^2}{2}\omega^{kj}\partial_j\omega^{il}\partial_k\partial_l-\frac{\theta^ 2}{12\mu}\partial _{l}\left( \mu \omega ^{ij}\partial
_{j}\omega ^{lk}\right)\partial_k+O\left( \theta ^{3}\right).  \label{15a}
\end{equation}
The definitions (\ref{scalar}) and (\ref{15}) means that the coordinate
operators are self-adjoint with respect to the introduced scalar product: $%
\left\langle \hat{x}^{i}\varphi \right\vert \left. \psi \right\rangle
=\left\langle \varphi \right\vert \left. \hat{x}^{i}\psi \right\rangle $.
\textit{The momentum operators} $\hat{p}_{i}$ are fixed from the condition
that they also should be self-adjoint with respect to (\ref{scalar}). One of the
possibilities is to choose it in the form
\begin{equation}
\hat{p}_{i}=-i\partial _{i}-\frac{i}{2}\partial _{i}\ln \mu \left( x\right) .
\label{p}
\end{equation}%
One can easily verify that in this case $\left\langle \hat{p}_{i}\varphi
\right\vert \left. \psi \right\rangle =\left\langle \varphi \right\vert
\left. \hat{p}_{i}\psi \right\rangle .$

The momentum operators (\ref{p}) commute, $[\hat{p}_{i},\hat{p}_{j}]=0$. The commutator between $\hat{x}^{i}$, defined in (\ref{15a}), and $\hat{p}_{j}$ is
\begin{equation}
\left[ \hat{x}^{i},\hat{p}_{j}\right] =i\delta^i_j-\frac{i\theta}{2}\left( \partial_j\omega^{il}\left(\hat{x}%
\right)\hat{p}_{l}+ i\partial_j \left(\omega^{il}\partial_l\ln \mu\right) \left( \hat x\right) \right)+O\left( \theta ^{2}\right).  \label{xp}
\end{equation}
So, the complete algebra of commutation relations involving $\hat{x}^{i}$ and $\hat{p}_{j}$ is a deformation in $\theta$ of a standard Heisenberg algebra. Finally, it should be noted that all expansions in our approach are formal, we do not discuss the convergence of preturbative series here.

\section{Particle in a central potential}

Let us consider a particle placed in a central potential $V\left(r^2/2\right)$, where $r^{2}=x^{2}+y^{2}+z^{2}$. The
corresponding Hamiltonian reads
\begin{equation}
\hat{H}=\frac{\hat{p}^{2}}{2}+V\left( \frac{\hat{r}^{2}}{2}\right) .
\label{H}
\end{equation}
As it was already mentioned in the introduction, the canonical noncommutativity which can be realized by the coordinate operators $\hat{x}^{i}_{can}={x}^{i}+i/2\theta ^{ij}\partial_j$ violates the rotational symmetry of (\ref{H}), since $[L_{i},V(\hat{x}^2_{can}/2)]\neq0$, where $%
L^{i}=-i$ $\varepsilon ^{ijk}x_{j}\partial _{k}$ is the angular momentum operator. And this problem cannot be solved introducing a noncommutativity as a Drinfeld
twist deformation \cite{Toppan1}. See also \cite{Toppan2} for an example of
nonabelian twist in QM.

The external antisymmetric field $\omega ^{ij}\left( x\right) $ can be
chosen in a way to preserve a rotational symmetry of the system. Let us
suppose that
\begin{equation}
\omega ^{ij}=\varepsilon ^{ijk}x^{k}f\left( r^{2}\right) ,  \label{omega}
\end{equation}%
where $f$ is a given function. The corresponding noncommutative algebra is
\begin{equation}
\left[ \hat{x}^{i},\hat{x}^{j}\right] =i\theta\varepsilon ^{ijk}\hat{x}^{k}f\left( \hat{r}^{2}\right).\label{1a}
\end{equation}
The consistency condition for this algebra, $\left[ \hat{x}^{i},\varepsilon ^{jkl}\hat{x}^{l}f\left( \hat{r}^{2}\right)\right]+cycl.(ijk)=0,$ is satisfied automatically and
no corrections $\omega
_{co}^{ij}\left( x\right)$ are needed to the (\ref{omega}) to construct the explicit form of the star product in any order in $\theta$ using the iterative procedure \cite{KV}. Note that the algebra (\ref{1a}) is a generalization of the algebra of fuzzy sphere \cite{Fuzzy}. In particular, the rotationally invariant NC space can be obtained as a foliation of fuzzy spheres \cite{Hammou}. Moreover, since $\left[ \hat{x}^{i}, \hat{r}^{2}\right]=0,$
one can introduce new NC coordinates $\hat X_k = \hat x_k/f(\hat r^2)$
that satisfy simpler commutation relations $\left[\hat X_i, \hat X_j\right] =
i \theta \varepsilon_{ijk} \hat X_k$ and possess simple operator
realizations.

One can easily see that any function $\mu (r^{2})$ obeys the equation (\ref%
{4}):
\begin{equation*}
\partial _{i}\left( \mu (r^{2})\varepsilon ^{ijk}x^{k}f\left( r^{2}\right)
\right) =0,
\end{equation*}%
and can be chosen as a measure to define a trace functional. For simplicity
we set $\mu \left( x\right) =1$. So, the momentum operators are just
derivatives $\hat{p}_{i}=-i\partial _{i}$. According to the previous section these operators are self-adjoint with respect to the inner product (\ref{scalar}), defined on the configuration space $\textbf{R}^3$ with a measure $\mu \left( x\right) =1$. However, some QM models with a rotational invariant potentials, like e.g. the Hydrogen atom, are well defined on the configuration space $\textbf{R}^3_0=\textbf{R}^3\setminus\{0\}$. In this case the property of the momentum operator to be self-adjoint should be investigated separately, see e.g., \cite{Roy} for this discussion in a standard QM.

Taking into account the above definition of the momentum operators and the explicit form of the external antisymmetric field (\ref{omega}) the Hamiltonian (\ref{H}) takes the form:%
\begin{eqnarray}
&&\hat{H}=-\frac{1}{2}\Delta+V\star ^{\prime }= \\
&&-\frac{1}{2}\Delta+V-\frac{\theta
^{2}}{12}V^{\prime }f^{2}\left( r^{2}\Delta -2x^{i}\partial
_{i}-x^{i}x^{j}\partial _{i}\partial _{j}\right) +O\left( \theta ^{4}\right)
=\notag \\
&&-\frac{1}{2}\Delta+V+\frac{\theta ^{2}}{12}V^{\prime }f^{2}L^{2}+O\left( \theta ^{4}\right) ,
\notag
\end{eqnarray}%
where $L^{2}=L_{x}^{2}+L_{y}^{2}+L_{z}^{2}$ is the the orbital angular
momentum, and $V^{\prime }$ means the simple derivative of the function $V$ with respect to the argument. In particular, if $V=(r^2)^{-1/2}$, then $V^{\prime}=-\frac{1}{2}(r^2)^{-3/2}$. Since in Cartesian coordinates the integration measure is $\mu \left( x\right) =1$, writing the perturbed Hamiltonian and an internal product in the spherical coordinates one may easily verify that this Hamiltonian is self-adjoint. The specific choice of the
external field (\ref{omega}) implies the conservation of the rotational
symmetry:%
\begin{equation}
\left[ L_{i},V\left( \frac{1}{2}\hat{r}^{2}\right) \right] =0.
\end{equation}%
This, in turn, means the preservation of the degeneracy of the corresponding
energy spectrum over the magnetic quantum number $m$.

In particular, for the Coulomb potential the leading order perturbation due to
noncommutativity is%
\begin{equation}
\hat{V}_{NC}=-\frac{e^{2}\theta ^{2}}{12}\frac{f^{2}L^{2}}{r^{3}}.
\end{equation}%
We use the usual perturbation theory to calculate the leading corrections to
the energy levels,%
\begin{equation}
\Delta E_{n}^{NC}=\left\langle \psi ^{0}\left\vert \hat{V}_{NC}\right\vert
\psi ^{0}\right\rangle .
\end{equation}%
According to Messiah \cite{Messiah} we use the following expressions for the unperturbed normalized position wave function in spherical coordinates:
\begin{equation}
\left\vert \psi ^{0}\right\rangle =\sqrt{\left(\frac{2}{na_0}\right)^3\frac{(n-l-1)!}{2n(n+l)!}}e^{-r/na_0}\left(\frac{2r}{na_0}\right)^l
L^{2l+1}_{n-l-1}\left(\frac{2r}{na_0}\right)Y^m_l(\vartheta,\varphi),
\end{equation}
corresponding to the energy $%
E_{n}=-e^{2}/2a_{0}n^{2}$, here $n$ is a principal quantum number and $%
a_{0}=1/\alpha $ is the Bohr radius. Integrating over the angular variables $\vartheta$ and $\varphi$, we stay with%
\begin{equation}
\Delta E_{n}^{NC}=-\frac{e^{2}\theta ^{2}l(l+1)}{12}\left\langle \frac{f^{2}%
}{r^{3}}\right\rangle ,
\end{equation}%
where $l$ is the azimuthal quantum number.

In order to calculate the corresponding nonlocality we first write (\ref{1a})
as
\begin{equation}
\left[ \hat{x}^{i},\hat{x}^{j}\right] =i\theta f
\varepsilon ^{ijk}{x}^{k}+\frac{i\theta ^{2}}{2}f^2\varepsilon ^{ijk}{L}%
_{k}+O(\theta ^{3}).
\end{equation}%
Taking into account that $\left\langle \psi ^{0}\left\vert f\left( {r}%
^{2}\right) x^{k}\right\vert \psi ^{0}\right\rangle =0$, due to the
integration over the angular variables, we end up with
\begin{equation}
\Delta x\Delta y\geq \frac{\theta ^{2}}{4}\left|m\langle f^{2}\rangle\right| ,
\end{equation}%
where $m$ is a magnetic quantum number.

We will consider two possibilities for the function $f$ here: $f=1$ and $%
f=r$. In the both cases the external field is zero at the origin and then grows along the corresponding axis, however in the second case the growth is higher. In the first case, taking into account the fact that%
\begin{equation*}
\left\langle \frac{1}{r^{3}}\right\rangle =\frac{1}{n^{3}a_{0}^{3}}\frac{1}{%
l\left( l+1/2\right) \left( l+1\right) },
\end{equation*}%
the energy level shift is
\begin{equation}
\Delta E_{n}^{NC}=-\frac{\theta^2 E_n^2}{3 a_0e^2}\frac{n}{l+1/2}.
\end{equation}%
This energy correction has the same form as a relativistic kinetic energy
contribution to the fine structure. Nonlocality in this case is defined by the relation
\begin{equation}
\Delta x\Delta y\geq \frac{\theta ^{2}}{4}|m|.
\end{equation}

Reminding that $\left\langle \frac{1}{r}\right\rangle =1/n^2a_o$, the energy correction for the second possibility $(f=r)$ is given by
\begin{equation}
\Delta E_{n}^{NC}=\frac{\theta^2}{6}E_nl(l+1).
\end{equation}%
Finally since,
\begin{equation*}
\left\langle r^2\right\rangle =\frac{n^2a_0^2}{4}\left(4n^2-l^2+2nl+1\right),
\end{equation*}%
the corresponding nonlocality is given by
\begin{equation}
\Delta x\Delta y\geq |m|\frac{\theta ^{2}n^2a_0^2}{16}\left(4n^2-l^2+2nl+1\right).
\end{equation}
We can see that in this case the nonlocality depends on the energy of the system, the more the energy, the more the nonlocality.

The alternative approach to the Hydrogen atom on rotationally invariant NC space, obtained as a sequence of fuzzy spheres \cite{Hammou}, was discussed in \cite{Presnajder}. In this paper the Hamiltonian of the system was constructed in such a way that all the symmetries of H-atom were conserved, preserving the degeneracy of the energy spectrum over the magnetic $m$ and the azimutal $l$ quantum numbers. While in our construction only the rotational symmetry of the system remains untouchable, but the degeneracy over the azimutal quantum number $l$ is removed. We see the advantage of our approach to NCQM in the possibility of generalization. It is applicable to the case of the arbitrary NC space (\ref{1}), see the Sec. 2, and not only to the fuzzy spaces. However, our approach is essentially perturbative construction and it does not allow to recover non-perturbative consequences of the underlying noncommutative space. To study such a consequences one will need an explicit form of the star product on the considered NC space, which can be found in \cite{Presnajder1,Pinzul}.

\section{Conclusions and perspectives}

This example shows that the noncommutativity can be introduced in a minimal way in the theory, i.e.,
one may obtain nonlocality without violating physical observables like the energy
spectrum, etc.

The considered model also indicates the way how to construct a relativistic generalization of QM with NC coordinates \cite{kup14}. If the external field $\omega^{\rho\sigma}(x)$ which determine the commutation relations $\left[ \hat{x}^{\rho },\hat{x}^{\sigma }\right] =i\theta \hat{\omega}^{\rho\sigma
}\left( \hat{x}\right) ,$ transforms as a two tensor with respect to a Lorentz group, the operators $\hat{x}^{\rho }=x^{\rho }+i\theta /2\omega ^{\rho\sigma }\partial
_{\sigma }+O\left( \theta ^{2}\right) $ and $\hat{p}_{\rho }=-i\partial _{\rho
}-i\partial _{\rho }\mu \left( x\right) $ will transform as vectors. This fact may be used to construct relativistic wave equations on curved NC space-time. For simplicity one may start with $(2+1)$ dimensions, choosing the external antisymmetric field as $\omega^{\rho\sigma}(x)=\varepsilon^{\rho\sigma\lambda}x_\lambda f(x^2)$.

Alternative models of relativistic QM with noncommutative coordinates were constructed in \cite{GKS} in $(3+1)$ dimensions and in \cite{Gamboa1} in $(2+1)$ dimensions  as a relativistic version of spin noncommutativity \cite{Gamboa2}. However, in this case the commutator between coordinates is proportional to the spin, i.e., coordinates do not form an algebra. This lead to the difficulties in the definition of the star product and makes the operators to be non Hermitian \cite{FGKS}.


\begin{thebibliography}{9}
\bibitem{Doplicher} S. Doplicher, K. Fredenhagen and J. Roberts,
Commun.Math.Phys. \textbf{172} (1995) 187.

\bibitem{KV} V.G. Kupriyanov, D.V. Vassilevich, Eur.Phys.J.C. \textbf{58}
(2008) 627-637.

\bibitem{GK} M. Gomes, V.G. Kupriyanov, Phys.Rev.\textbf{D79} (2009) 125011.

\bibitem{Fring1} A. Fring, L. Gouba, F.G. Scholtz,  J.Phys. A43 (2010) 345401.

\bibitem{Fling2} B. Bagchi, A. Fring, Phys. Lett. A373 (2009) 4307.

\bibitem{Meljanac} E. Harikumar, T. Juric, S. Meljanac, Phys.Rev.\textbf{D84} (2011) 085020; Phys.Rev.\textbf{D86} (2012) 045002.

\bibitem{Presnajder} V. Galikova, P. Presnajder, J. Phys: Conf. Ser. 343 (2012) 012096.

\bibitem{kup14} V.G. Kupriyanov, \textit{General form of quantum mechanics
with noncommutative coordenates,} arXiv:1204.4823 [math-ph].

\bibitem{Chaichian} M.~Chaichian, M.M.~Sheikh-Jabbari and A.~Tureanu,
Phys.Rev.Lett.\ \textbf{86} (2001) 2716.

\bibitem{AGSV} T.C. Adorno, D.M. Gitman, A.E. Shabad, D.V. Vassilevich,
Phys.Rev.\textbf{D84} (2011) 085031.

\bibitem{Kontsevich} M.~Kontsevich, \ Lett.\ Math.\ Phys.\ \textbf{66} (2003) 157.

\bibitem{FS} G. Felder, B. Shoikhet, Lett. Math. Phys. \textbf{53} (2000)
75 .

\bibitem{Toppan1} B. Chakraborty, Z. Kuznetsova, F. Toppan, J.Math.Phys. 51
(2010) 112102.

\bibitem{Toppan2} P.G. Castro, R. Kullock, F. Toppan,
J.Math.Phys. 52 (2011) 062105.

\bibitem{Fuzzy} J. Madore, Class.Quant.Grav. 9 (1992) 69; H.Grosse, J.Madore, H.Steinacker, Int.J.Mod.Phys. A {\bf 17} (2002) 2095.

\bibitem{Hammou} A.B.Hammou, M.Lagraa and M.M.Sheikh-Jabbari, Phys. Rev.D
{\bf 66} (2002) 025025; E.Moreno, Phys. Rev. D {\bf 72} (2005) 045001.

\bibitem{Roy} U.Roy, S.Gosh, K.Bhattacharya, Revista Mexicana de Fisica \textbf{E 54} (2) (2008) 160.

\bibitem{Messiah} Messiah, Albert (1999). Quantum Mechanics. New York: Dover.

\bibitem{Presnajder1} P. Presnajder, J. Math. Phys. 41(5) (2000) 2789.

\bibitem{Pinzul} G. Alexanian, A. Pinzul, A. Stern, Nucl.Phys. \textbf{B600} (2001) 531.

\bibitem{GKS} M. Gomes, V.G. Kupriyanov, A.J. da Silva, Phys.Rev.\textbf{D81} (2010) 085024.

\bibitem{Gamboa1} H.Falomir, F.Vega, J.Gamboa, F.Mendez, M.Loewe, Phys.Rev. \textbf{D86} (2012) 105035.

\bibitem{Gamboa2} H. Falomir, J. Gamboa, J. Lopez-Sarrion, F. Mendez, P.A.G. Pisani, Phys.Lett \textbf{B680} (2009) 384.

\bibitem{FGKS} A.F. Ferrari, M. Gomes, V.G. Kupriyanov, C.A. Stechhahn,  Phys.Lett. \textbf{B718} (2013) 1475.

\end{thebibliography}
\end{document}